\documentclass[twocolumn,reprint,aps,prb,showkeys,showpacs,groupaddress,superscriptaddress,sectionbib,square]{revtex4-1}
\setcitestyle{numbers,square}
\usepackage{hyperref}
\usepackage{epsfig}
\usepackage{graphicx}
\begin{document}
\title{Atomic motion in solids with dimpled potentials} 
\author{N.~A. Zarkevich}  
\email{zarkev@ameslab.gov} 
\affiliation{The Ames Laboratory, U.S. Department of Energy, Ames, Iowa 50011-3020 USA}

\date{\today}
\begin{abstract} 
Polymorphic solids of the same chemical composition can have different atomic structures;  in each polymorph atoms vibrate around a local potential energy minimum (LPEM). If transformations to other structures have sufficiently high enthalpy barriers, then each polymorph is either stable or metastable; it is stationary and does not spontaneously change with time. But what happens, if those barriers are low? 
As examples, we consider NiTi shape memory alloy exhibiting a large elasto\-caloric effect, and selected elemental solids. 
We suggest a model for dynamically polymorphic solids, where multiple LPEMs are visited by ergodic motion of a single atom.  We predict that upon cooling a dynamically polymorphic phase should undergo a symmetry-breaking first-order phase transition, accompanied by a finite change of the lattice entropy. 
We discuss 3 methods used to calculate phonons in solids with non-harmonic dimpled atomic potentials, and compare theoretical predictions to experiment. 
\end{abstract}


\maketitle

\section{\label{sIntroduction}Introduction}
A non-harmonic atomic potential presents a challenge for those who use a harmonic or quasi\-harmonic approximation for addressing solids.  
A dimpled potential is not harmonic, can cause a lattice instability, is not straightforward to deal with,  but is very common in practical materials \cite{RMP84p945y2012}  
with 
more than one 
local potential energy minimum (LPEM), covered by ergodic atomic motion. 
There are well-developed methods for a harmonic potential, where parabola has a single minimum.  
However, a dimpled potential dramatically differs from a harmonic one.  
Quite often, the high-symmetry atomic position, which used to be an energy minimum in a harmonic potential,
happen to be a local energy maximum or a saddle in a dimpled potential, with several LPEMs surrounding this position. 
How to calculate phonons and describe atomic motion in such non-harmonic potentials?  
We will consider 3 algorithms, obtain \emph{ab initio} results, and compare them to experiment. 

{\par}
	Crystal is a solid, in which atoms are arranged in a definite pattern and whose surface regularity reflects its internal symmetry \cite{britannicaCrystal}. 
A crystal can be described as a Bravais lattice -- an infinite periodic array of discrete points \cite{Ashcroft1976}.   
The crystalline periodic arrangement of atoms is manifested by an x-ray \cite{Bragg1913} and neutron diffraction crystallography. 
According to Laue \cite{Laue1915Nobel}, the Bragg spots \cite{Bragg1913} are observed 
due to constructive interference, which occurs if the change in the wave vector belongs to the reciprocal lattice. 
Nevertheless, in some crystals the ideal atomic positions on a Bravais lattice are unstable \cite{RMP84p945y2012}. 

{\par}
	Comparing an experimentally observed diffraction pattern with the one predicted for a particular Bravais lattice, crystallographers suggest a crystal structure \cite{Bragg1913}.  
However, a correspondence between crystal structures and diffraction patterns is many-to-one: more than one crystal structure can produce the same pattern, while each particular fully ordered crystal produces a unique (one and only one) diffraction pattern.  To add confusion, a partially disordered crystal can produce a similar pattern.   
In addition to thermal atomic motion, materials can have athermal atomic disorder, 
which can be chemical 
or displacive, 
local (due to lattice defects) or non-local.  
An ordered crystal and a solid with a substantial 
displacive disorder
can produce similar diffraction patterns  
with the same positions but different broadening of  peaks. 
One example is NiTi austenite 
with the assumed unstable B2 structure and multiple stable \emph{representative} structures \cite{PRB90p060102,PRL113p265701}.  
Lattice instabilities were also found in antiferromagnetic (AFM) phase of B2 FeRh \cite{PRB97p014202y2018} and in body-centered cubic (bcc) phases of Ti, Zr, Hf, and Li \cite{PRL53p64y1984,PRB43p10933y1991,RMP84p945y2012}. 
A crystal-like diffraction pattern can be produced by solids, which are not periodic (for example, quasicrystals)  \cite{quasicrystals}. 

\begin{figure}[b]
\includegraphics[width=80mm]{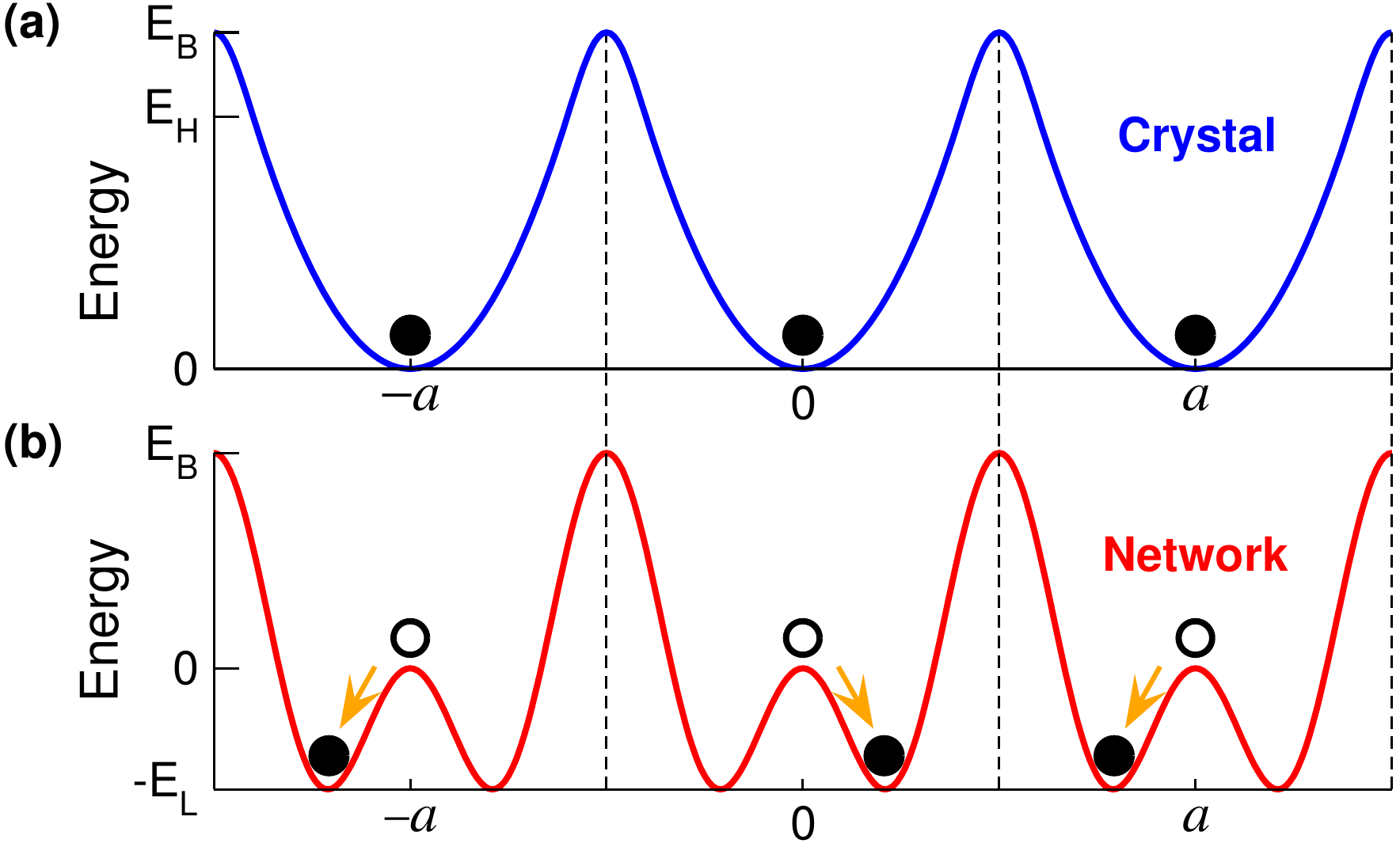} 
\vspace{-2mm}
\caption{\label{Fig1ab}(color online) 
A 1-dimensional periodic potential with the lattice constant $a$, having 
(a) single LPEM (harmonic below $E_H$), 
and 
(b) multiple (two) LPEMs per basin, where 
at $kT \ll E_L$  each atom is displaced (orange arrows) from the high-symmetry unstable position (open circles) 
to one of the nearby stable LPEMs (filled circles).  
}
\end{figure}

{\par}
	Atoms in solids are trapped in deep potential energy (PE) basins (Fig.~\ref{Fig1ab}), separated by PE barriers $E_B$, which are high compared to $kT$, 
where $T$ is temperature and $k$ is the Boltzmann constant.
We define $N_L$ to be the number of the local potential energy minima 
per PE basin. 
We refer to a crystal with $N_L=1$ as conventional; a 1-dimensional (1D) example 
is in Fig.~\ref{Fig1ab}(a). 

{\par}
A solid with a dimpled potential can have multiple LPEMs and hence $N_L>1$  [e.g., $N_L=2$ in Fig.~\ref{Fig1ab}(b)].  
In general, potential energy is a functional of atomic positions, and a path from one LPEM to another can be a collective atomic motion. 

{\par}
Each LPEM is a stable or metastable arrangement of atoms.  Multiple stable chemical structures are known as isomers in molecules and polymorphs in solids. 
If the enthalpy barriers between polymorphs are high compared to $kT$, then each 
polymorph is stationary and does not spontaneously transform to other polymorphs. 
However, if the barriers between LPEMs are low compared to $kT$, then thermal atomic motion covers several LPEMs;  
this state of matter can be called ``dynamically polymorphic'', 
and this phenomenon -- 
a ``dynamic polymorphism.'' (A similar semantics is used in computer programming 
to refer to the runtime polymorphism). 

{\par}
	Below we discuss atomic motion in a dimpled potential (section~\ref{sGeneric}) and compare phonon methods (section~\ref{sPhonon}). 
	As a model, we consider a periodic potential with deep PE basins; only one atom occupies each basin 
(we assume that a strong inter\-atomic repuslion at short distances makes presence of another atom in the same basin energetically unfavorable).  
In our examples (section~\ref{Examples}) we consider one [Figs.~\ref{Fig1ab}(a) and 2] or more [from two in Fig.~\ref{Fig1ab}(b) to 48 in Fig.~\ref{Fig48cub}(c,d)] LPEMs per basin.   
Properties of the simplified models and real materials are discussed in sections~\ref{sGeneric}, \ref{Examples}, \ref{sDiscussion} and summarized in section~\ref{sConclusion}.   
Computational details are provided in Appendix~\ref{Acompute}.

\section{\label{sGeneric} Dimpled atomic potential}
\subsection{\label{aModel}Model}

{\par}
Let us consider a solid with several $(N_L\!>\!1)$ LPEMs per PE basin. 
	In a single PE basin, a set of LPEMs connected by minimal enthalpy paths (MEPs) forms a network, 
which might include (Figs.~\ref{Fig1ab}(b) and \ref{Fig3doublewell}) or exclude (Figs.~\ref{Fig4pot2D}, \ref{Fig5Corrugated} and \ref{Fig48cub})
the high-symmetry crystallographic position at the basin center ($x=0$, with energy $E_L$ above LPEM). 
Let the enthalpy barriers along MEPs be $E_l$ (relative to LPEM); 
for simplicity we assume that enthalpies of all LPEMs are the same and the barriers have the same height 
(one can index individual enthalpies and generalize our consideration to less symmetric cases).  
At a high enough temperature  ($kT \ge E_l$), an atom can move from one LPEM to another within the same basin. 

{\par} 
	We assume that the enthalpy barrier $E_l$ is not higher than the PE $E_L$ at the high-symmetry point $x=0$, 
and both enthalpies are low compared to the barrier $E_B$ between the PE basins, see Fig.~\ref{Fig1ab}(b):
\begin{equation}
\label{ElEL}
0 \le  E_l \le E_L \ll E_B .
\end{equation}  
	In our examples, 
$E_l =  E_L =0$ for the harmonic potential with $N_L =1$ in Figs.~\ref{Fig1ab}(a) and \ref{Fig2harm}; 
$E_l \equiv  E_L >0$ for the 
 double-well potential with $N_L =2$ in Figs.~\ref{Fig1ab}(b) and \ref{Fig3doublewell}; 
 $0 = E_l < E_L$ for the muffin-tin sombrero potential with $N_L = \infty$ in Figs.~\ref{Fig4pot2D} and \ref{Fig2D2}; 
 $0 < E_l \ll E_L$ for the 2D and 3D corrugated sombrero potentials with a finite countable $N_L>1$ in  Figs.~\ref{Fig5Corrugated} and \ref{Fig48cub}. 

{\par} 
	In general, one of the LPEMs can be at $x=0$ (for example, the only LPEM  in a harmonic potential in Fig.~\ref{Fig2harm}). 
Here we focus on a less trivial atomic motion with an instability at $x=0$; the MEP between adjacent stable LPEMs 
can either include 
or bypass the high-symmetry point $x=0$. 
	The double-well basin in Fig.~\ref{Fig3doublewell} is a 1D example of a potential with $N_L=2$ [Fig.~\ref{Fig1ab}(b)],   
where the path from $(-x_L, -E_L)$ to $(+x_L, -E_L)$ unavoidably goes through the local PE maximum at $(0, 0)$,  
thus $E_l \equiv E_L$.   
In higher dimensions $D \! > \! 1$, a MEP can go around this local maximum (examples are 2D and 3D corrugated sombreros in Figs.~\ref{Fig5Corrugated} and \ref{Fig48cub}). 

{\par}  Assuming that each LPEM is a point (and not a line, like in the muffin-tin sombrero potential in Fig.~\ref{Fig4pot2D}), 
one can define a harmonic limit $E_h < E_l$ around it, 
such that atomic vibration around a single LPEM is harmonic at sufficiently small displacements 
and potential is approximately parabolic at $E \le E_h$.

\subsection{Thermal atomic motion}
{\par} 
	Atomic motion in a solid with $N_L >1$ depends on $T$
and can  cover vicinity of one or several LPEMs. 
\begin{itemize}
\item
$kT \le E_h \ll E_l$:  harmonic vibration around a single local potential energy minimum. 
The small atomic displacement method can be used to calculate phonons at a LPEM, see Section \ref{sPhonon}.
\item
$E_h<kT<E_l$: anharmonic vibration around a single LPEM.
\item
$kT \sim E_l$: a phase transition happens. 
\item
$E_l \le kT < E_L$:  atomic motion covers several LPEMs in the same PE basin. 
If such LPEMs are distributed symmetrically around $x=0$, 
then the time-averaged atomic position is zero: $\left< x \right> =0$.  
\item
$E_L \le kT<E_B$:   atomic motion covers a significant part of the PE basin, including the center ($x \! = \! 0$) and multiple LPEMs. 
If LPEMs can be interpreted as a negligible roughness ($E_L \ll kT$) at the bottom of a nearly harmonic potential,   
then a finite atomic displacement method can be used to calculate phonons around $x=0$, see Section~\ref{sPhonon}. 
Snapshots of thermal atomic motion in molecular dynamics at fixed $T$ are finite collective atomic displacements in a solid phase. 
\item
$kT \ge E_B$: atomic motion is no longer restricted by a PE basin; the solid has melted or sublimated. 
\end{itemize}

{\par}
	A dynamically polymorphic solid phase exists at temperatures $E_l \le  kT < E_B$. 
Upon cooling, it transforms to a lower-symmetry phase below 
\begin{equation}
\label{Tc}
   T_c \approx E_l/k .
\end{equation}  
We expect this transformation to be of the first order, 
because it is accompanied by a discontinuous change of the lattice entropy $\Delta S_L$.  
Under certain  conditions, a change of the total entropy is responsible for an isentropic temperature change (caloric effect). 
In alloys like NiTi, this effect is quite large.

\subsection{\label{Representative}Representative structures}
{\par}  
	At $kT \ll E_l$, atomic arrangement is stationary:  each atom vibrates around a single LPEM.
If at each PE basin an atom  randomly chooses one of the LPEMs 
[e.g., either right or left LPEM in each double-well basin in Fig.~\ref{Fig1ab}(b)], 
then this atomic structure is aperiodic, but its diffraction pattern coincides with that produced by a crystal with partially occupied LPEMs in periodic PE basins. 
Even if there is no atomic periodicity, 
it is possible to consider a ``large enough'' representative periodic unit cell, 
which correctly represents energy, phonon spectrum, average LPEM occupations,  
and other physical properties. 

{\par} 
Let us consider a 1D example in Fig.~\ref{Fig1ab}(b). 
A periodic unit cell with even number of atoms, where each (right or left) LPEM is occupied by half of the atoms, 
 provides the same energy per atom (at $T=0$), 
the same occupancy ($c_i = 1/2$) of each (right or left) LPEM, 
and a vibrational spectrum similar to that of the whole aperiodic solid. 
Thus,  for the purpose of computing its physical properties,  a solid can be approximated by a periodic atomic configuration with a  \emph{representative} structure.  
Any particular representative unit cell is not unique, there are many others.  All \emph{representative} structures have similar properties, which approximate those of the solid. 

\subsection{\label{DisplacivePattern}Displacive pattern}
{\par} 
Displacements of atoms from the high-symmetry positions (PE basin centers at $x=0$)  create a pattern, 
which can be either stationary (at $kT < E_l$) or dynamic (at $E_l \le  kT < E_B$), 
and has a characteristic length \cite{JETP40n1p1y1974}, imported from cosmology \cite{Kibble1980} to condensed matter \cite{Zurek1996}. 

{\par} 
	At $kT \ge E_l$, motion of each atom covers several LPEMs and the displacive pattern is dynamic. 
	A broadened distribution of the inter\-atomic distances 
differs from that in a conventional crystal (e.g., NiTi austenite \cite{PRB90p060102} has such broadening due to athermal atomic displacements).  
Although there is no periodicity of the instantly occupied LPEMs, a diffraction pattern produced by such a solid reminds that of a conventional crystal (with peaks at the same positions, but not of the same width).

\subsection{\label{InteratomicInteractions}Effect of inter\-atomic interactions}
{\par}
	Interactions between atoms can change relative energies and positions of LPEMs. 
A shift of energies can force atoms to choose one particular LPEM at each basin, thus forming a fully ordered crystal 
(an example is the $\gamma$-Se, hP3, A8, $P3_{1}21$ structure of Te and Se-Te alloys, 
which consists of 3 triangular lattices stacked along $z$ \cite{Hulin1963}, 
and can be considered as a distortion of a simple cubic lattice \cite{JETP59n6p1336y1984}).  
A shift of the equilibrium atomic positions affects a distribution of the inter\-atomic distances and a diffraction pattern. 
Nevertheless, atomic displacements and inter\-atomic distances in a dynamically polymorphic phase will differ from thermal ones in a harmonic crystal.

\section{\label{sPhonon} Phonon calculations}
\subsection{\label{small} Small atomic displacement method}
{\par}
	The quasi-harmonic approximation (QHA) is often used to calculate phonons in conventional crystals \cite{Kittel8}.  
Atoms are \emph{assumed} to be at stable equilibrium at 0K. 
Small atomic displacements $u$ from this stable equilibrium result in the increase in the potential energy $E$, 
which is quantified in QHA using a Taylor expansion: 
\begin{equation}
\label{U1}
	E = E_0 + \frac{1}{2} \sum_{ij, \alpha \beta} u_{\alpha} (r_i) D_{ij}^{\alpha \beta} u_{\beta} (r_j) + O (u_{m}^3) .
\end{equation}
Here $ D_{ij}^{\alpha \beta} $ is the force-constant matrix:
\begin{equation}
\label{D2}
   D_{ij}^{\alpha \beta} = \partial^2 E / \partial u_{\alpha} (r_i) \partial u_{\beta} (r_j) .
\end{equation}
The latin indices $i,j$ enumerate atoms, while the greek letters $\alpha, \beta$ label directions. 

	The higher-order terms beyond the second order can be neglected, if the amplitudes of all atomic displacements are small: $|u| \le u_{h}$, where $u_h$ is the harmonic limit.  
The instant atomic forces for the near-equilibrium atomic configuration $\tau_n$ are
\begin{equation}
\label{F3}
   F_i^{\alpha } (\tau_n) = \sum_{\beta, j} D_{ij}^{\alpha \beta} u^j_{\beta} (\tau_n)
\end{equation}
Given a sufficient number $N$ of independent atomic configurations $\tau_n$ ($n=1...N$) with known atomic displacements $ u^j $ and forces $ F_i $, 
one can solve a system of linear equations (\ref{F3}) and find matrix $ D_{ij} $, which can be used to find the phonon spectrum and density of states (DOS).  
The minimal number $N$ of independent atomic configurations [and linearly independent equations (\ref{F3})] 
is equal to the number $N_D$ of independent components of  $ D_{ij} $. 
The system (\ref{F3}) might be over-determined, if $N > N_D$.  
For either well-determined or over-determined system  (\ref{F3}), 
the effective force-constant matrix $D_{ij}^{e}$ can be found by  minimizing the sum of the differences between the actual and predicted forces \cite{ThermoPhonon}:
\begin{equation}
\label{F4}
    \Delta_F \equiv  \sum_{n,i} | F_i (\tau_n) -  \sum_j D_{ij}^{e} u^j (\tau_n) | \to \min
\end{equation}

{\par}
	Within the small atomic displacement method, expansion is around a LPEM, 
and displacement is \emph{assumed} to be within a harmonic limit (e.g., $|u| \le x_H$ in Fig.~\ref{Fig2harm}). 
Any displacement $|u|=|x-0| \le x_H$ (including infinitesimal) provides the same vibrational frequency for the harmonic potential in Fig.~\ref{Fig2harm}, where $ F(x)/x = dF/dx = K$ is a constant at $-x_H< x<x_H$. 
To emphasize that harmonicity is just an approximation, 
we exaggerated the harmonic region in Fig.~\ref{Fig2harm}, where 
the force $F(x)=Kx$ is precisely linear and the energy $E(x)=\frac{1}{2}Kx^2$ is parabolic at $x<x_H$.  

{\par}
	The result of the QHA is correct, 
if each displacement   $u=x-x_L$  from a stable equilibrium at $x_L$ is indeed within the harmonic limit at $|u|< u_h$, 
and
the potential energy $E(u)$ is approximately quadratic at $E< E_h \equiv E(x_L \pm u_h)$. 
To get a correct result, one must know the stable equilibrium coordinates $x_L$ at each occupied LPEM (Figs.~\ref{Fig1ab}(b) and \ref{Fig3doublewell}).

\subsection{\label{finite} Finite atomic displacement method}
\begin{figure}[t]
\includegraphics[width=80mm]{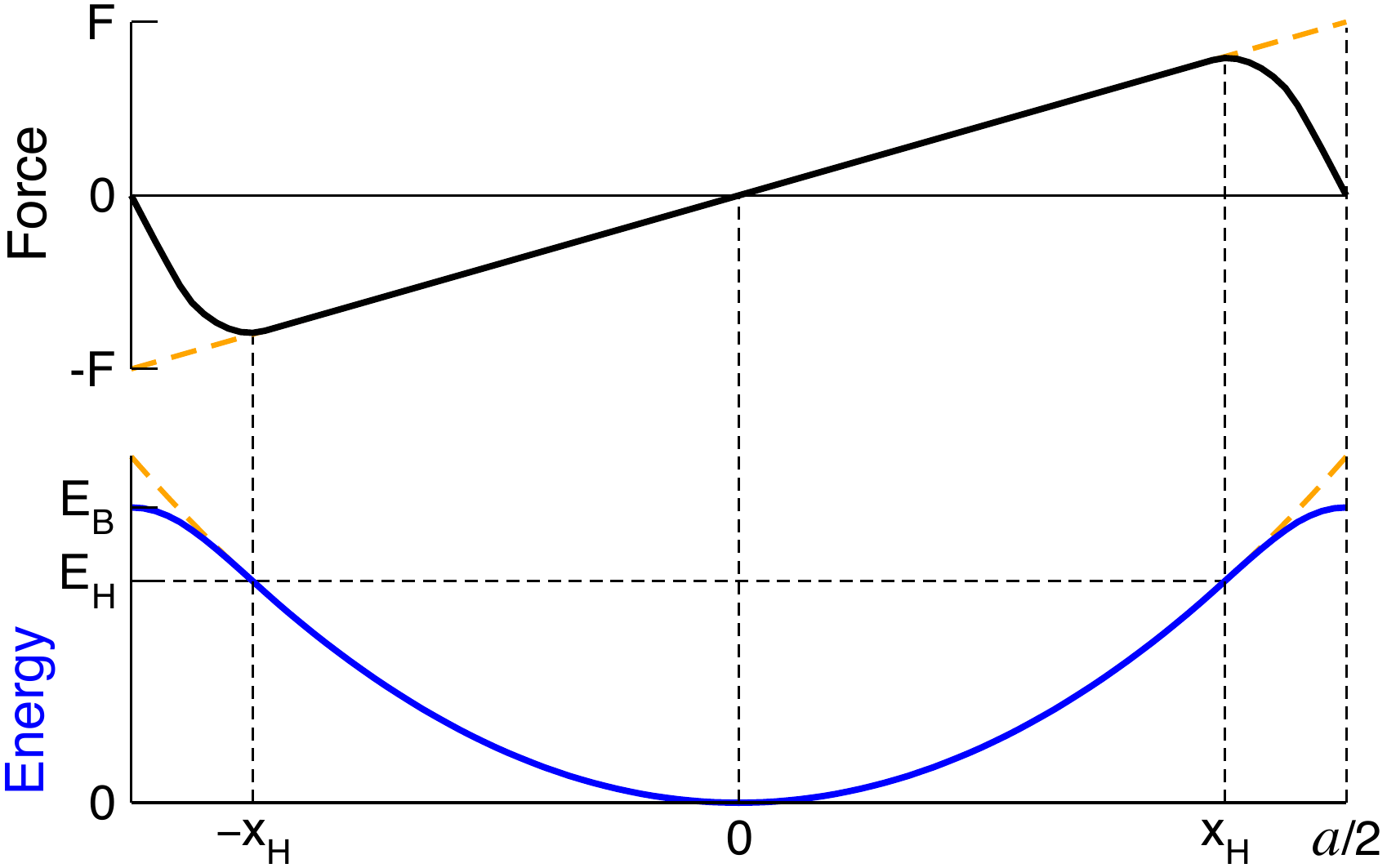} 
\vspace{-2mm}
\caption{\label{Fig2harm}(color online) 
Potential energy $E$ 
and force $F\!=\!dE/dx$ versus displacement $x$ (in arbitrary units) 
for a 1D crystal with one LPEM per basin [Fig.~\ref{Fig1ab}(a)], 
harmonic at $E\! <\! E_H$.  
}
\end{figure}

\begin{figure}[t]
\includegraphics[width=80mm]{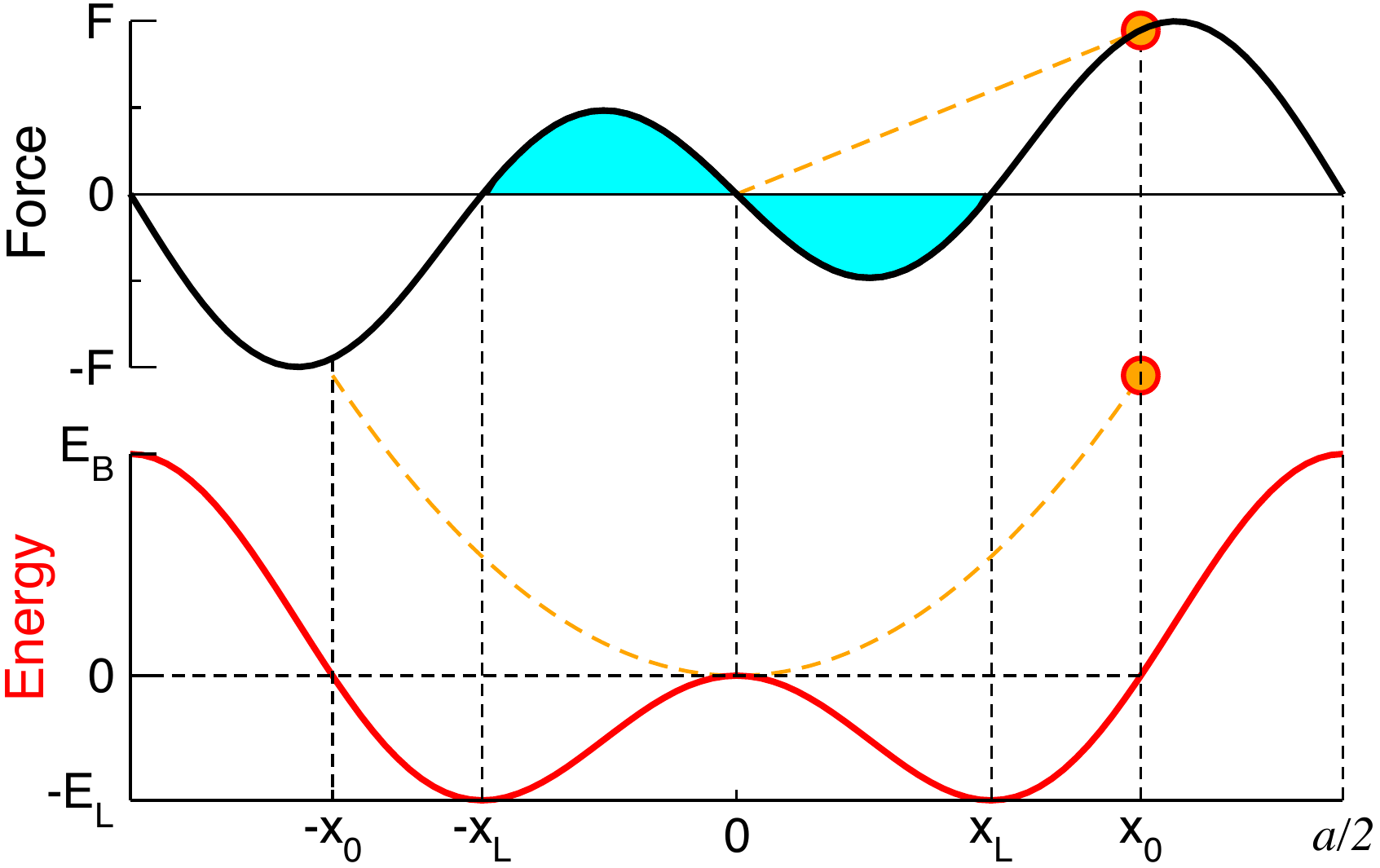}  
\vspace{-2mm}
\caption{\label{Fig3doublewell}(color online) 
Potential energy $E$ [same as in Fig. ~\ref{Fig1ab}(b)] and force $F=dE/dx$ versus displacement $x$ 
for a potential with several (two) LPEMs per potential energy basin.  
Highlighted are forces with $F/x\!<\!0$ at $-x_L\!<\!x\!<\!x_L$, which push an atom away from the unstable equilibrium at $x\!=\!0$.  
At displacement $x_0$ (filled circle), 
\emph{effective} harmonic linear force and parabolic potential are shown (dashed orange lines).
}
\end{figure}

{\par}
	Alternatively, the harmonic potential in Fig.~\ref{Fig2harm} can be comprehended as a limit of the double-well potential in Fig.~\ref{Fig3doublewell} 
with $x_L \to 0$, so that $x_0 = x_L = 0$ and $E(x_L) = E(0) = 0$. 
If multiple LPEMs can be interpreted as a negligibly small roughness at the bottom of a nearly harmonic potential (approximately parabolic at $-x_H < x < x_H$, with $x_L \to 0$, i.e., $x_L \ll  x_H$), 
then one can approximately calculate phonons using the finite displacement method (example is bcc Li, section~\ref{sbccLi}), 
which avoids unstable phonons, if $F(x)/x >0$;
 this happens when atomic displacements $u = x-0$ are larger than the distance between unstable (at zero) and stable (at $x_L$) atomic positions, see Fig.~\ref{Fig3doublewell}].

{\par}
	In the finite atomic displacement method, expansion (\ref{U1}) can be around the high-symmetry crystallographic position at $x=0$, 
which might (Fig.~\ref{Fig2harm}) or might not (Fig.~\ref{Fig3doublewell}) be a LPEM.   
If the finite displacements $u \equiv x-0$ are sufficiently large, i.e., $ x_L < | u | \le x_H $ and 
none of them is in the region $-x_L < x < x_L$, where a destabilizing force pushes an atom away from unstable equilibrium at $x=0$ (see Fig.~\ref{Fig3doublewell}), then there are no imaginary frequencies in the calculated phonon spectrum.  
Because this method deliberately avoids the region $-x_L < x < x_L$, unstable phonons are overlooked \cite{PRL100p095901}.

\subsection{\label{finiteT} Finite displacements at fixed T}
{\par}
	Molecular dynamics (MD) provides another way to generate atomic configurations $\tau (t)$.  
The snapshots of thermal atomic motion in MD at fixed $T$ are collective finite displacements. 
MD sampling gives a set of atomic positions $u^j (t_n)$ and forces $F_i (t_n)$ 
for a  large number $N$ of time steps $t_n$.  
ThermoPhonon code \cite{ThermoPhonon} solves the over-determined set of equations (\ref{F3}) and finds the effective force constant matrix $D_{ij}^{e}$ (\ref{F4}), 
which is then used 
to construct the phonon spectrum.


\section{\label{Examples}Examples}

\subsection{\label{s1Dharmonic} Harmonic potential}
{\par}
	Any smooth curve is approximately parabolic near its local minimum; 
it can be expanded in a Taylor series around it: $E(x)\approx E(0) + \frac{1}{2} K x^2 + O(x^3)$. 
Hence, a vibration with a small enough amplitude around a LPEM is expected to be harmonic 
(see Figs.~\ref{Fig1ab}(a) and \ref{Fig2harm} and discussion in section~\ref{small}).   
Average amplitude of atomic vibrations depends on T, and so does the harmonicity. 

{\par}
	For the potential in Fig.~\ref{Fig2harm} [same as Fig.~\ref{Fig1ab}(a)], one can categorize temperature dependence of atomic motion.
\begin{itemize}
\item[]$kT \le E_H$: harmonic vibration; 
\item[]$E_H<kT<E_B$: anharmonic vibration; 
\item[]$T_{melt} \sim E_B/k$: phase transformation \cite{PRL100p040602} from a solid to a liquid or a gas at $kT \ge E_B$. 
\end{itemize}

{\par}
	In one dimension, a $1 \! \times \! 1$ matrix $D^{\alpha \beta}$ (\ref{D2}) has only one element $D^{11}\!=\!K$. 
For the harmonic potential 
\begin{equation}
\label{Eharmonic}
E(x) =\frac{1}{2}Kx^2 ,
\end{equation} 
the force obeys the Hooke's law 
\begin{equation}
\label{Fharmonic}
  \frac{dE}{dx} \equiv F(x) =Kx; 
\end{equation} 
it is linear vs. $x$ at any ``small'' displacement $|x|<x_H$. 
Motion of a mass $M$ in such potential is harmonic and has a frequency 
\begin{equation}
\label{Wharmonic}
\omega = \sqrt{K/M} , 
\end{equation} 
which does not depend on $x$ at $|x|<x_H$.
	The Thermo\-Phonon method \cite{ThermoPhonon} at $kT<E_H$ and a displacement method at $|x|<x_H$ provide the same frequency (\ref{Wharmonic}). 

\subsection{\label{s1Danharmonic} Double-well potential}
{\par}
	A dimpled potential in Figs.~\ref{Fig1ab}(b) and \ref{Fig3doublewell} is a less trivial example with two LPEMs per basin. 
It has a crystallographic high-symmetry point at $x=0$, but $E(0)=0$ is a local energy maximum, 
with two minima nearby at $E(\pm x_L)=-E_L$. 
The point $(x,E)=(0,0)$ is the energy barrier $E_l = E_L$ between those two LPEMs. 

{\par}
At $kT \ll E_l \le E_L$, one can apply the small-displacement method (section~\ref{small}) to one of the representative stable structures (section~\ref{Representative}) to find phonons. 
Alternatively, at $E_L \ll kT < E_B$, if multiple LPEMs could be interpreted as a negligibly small roughness at the bottom of an otherwise nearly harmonic potential (similar to that in Fig.~\ref{Fig2harm}), 
then one could use a finite displacement method  
to approximate phonons at the unstable atomic position at $x=0$, see section~\ref{sPhonon}~(B,C). 

{\par}
	The 1D matrix $D^e$ in eq.~\ref{F4} has dimension $1 \times 1$ (i.e., $\alpha = \beta =1$), and the value of its single element is $K^e$.
	For the double-well potential in Fig.~\ref{Fig3doublewell}, the ``effective'' spring stiffness $K^e =F(x)/x$ depends on displacement $x$. 
A negative $K^e$ for a small displacement $x<x_L$ results in imaginary frequency  $\omega^e = \sqrt{K^e/M}$, which characterizes an unstable phonon mode. 
The displacement $x_L$ with $F(x_L)=0$ gives $K^e =0$ and $\omega^e = 0$. 
A large displacement $x>x_L$ leads to a positive $K^e$ and a stable effective phonon frequency $\omega^e >0$, which depends on displacement $x$. 
In other words, the frequency $\omega^e $ of vibrations around an unstable equilibrium at $x=0$ is not well-defined.  
However, a choice of a finite  $x_T \ge x_0$, related to thermal motion of atoms at temperature $T$, such that 
\begin{equation}
\label{ET}
 E(x_T)=\frac{1}{2}kT  ,
\end{equation} 
can result in a phonon spectrum, which compares well with experiment at the same $T$. 
In eq.~\ref{ET} we assumed $E_L \ll kT$ and neglected the dimples in Fig.~\ref{Fig3doublewell}. 

{\par}
	The ThermoPhonon method \cite{ThermoPhonon} at small $kT<E_L$  returns $K^e<0$ and an unstable (imaginary) phonon frequency for the expansion (\ref{U1}) around unstable equilibrium at $x=0$. 
With increasing $T$, the amplitudes of the negative $K^e$ and of the unstable imaginary phonon frequency $\omega^e = \sqrt{K^e/M}$ become smaller, 
until at a sufficiently large $kT \ge E_L$ the effective $K^e$ becomes positive.  If $K^e \ge 0$, than $\omega^e \ge 0$ is real and the effective phonon mode looks stable. 
Temperature $T$ uniquely determines the phonon frequency $\omega^e (T)$ within this method \cite{ThermoPhonon}. 

\begin{figure}[t]
\includegraphics[width=80mm]{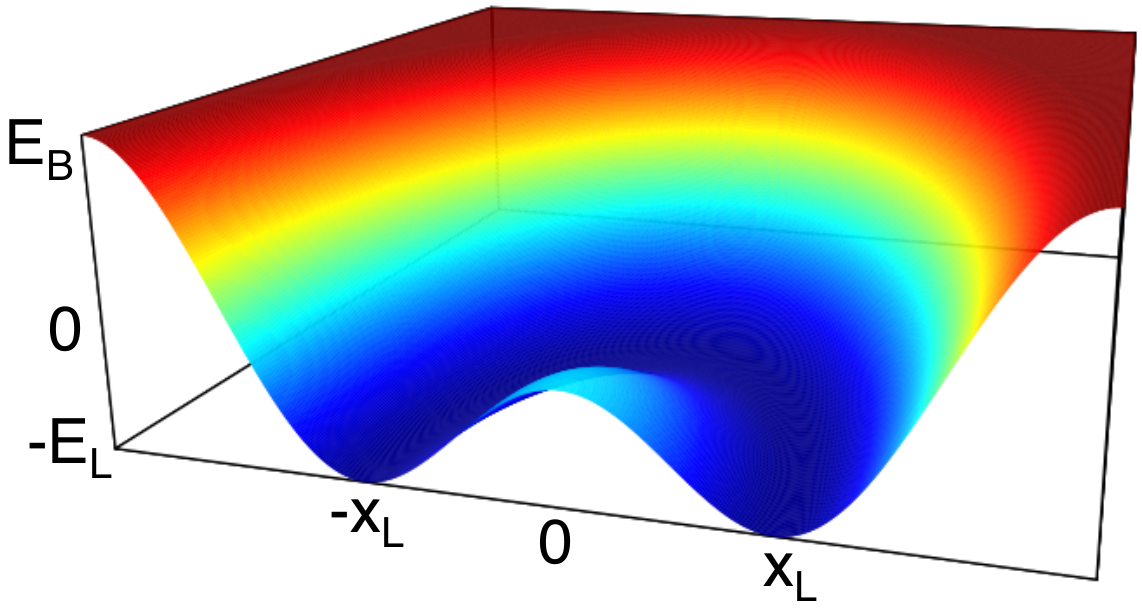}  
\vspace{-2mm}
\caption{\label{Fig4pot2D}(color online) 
The 2D sombrero-like model potential.  
}
\end{figure}
\begin{figure}[t]
\includegraphics[width=80mm]{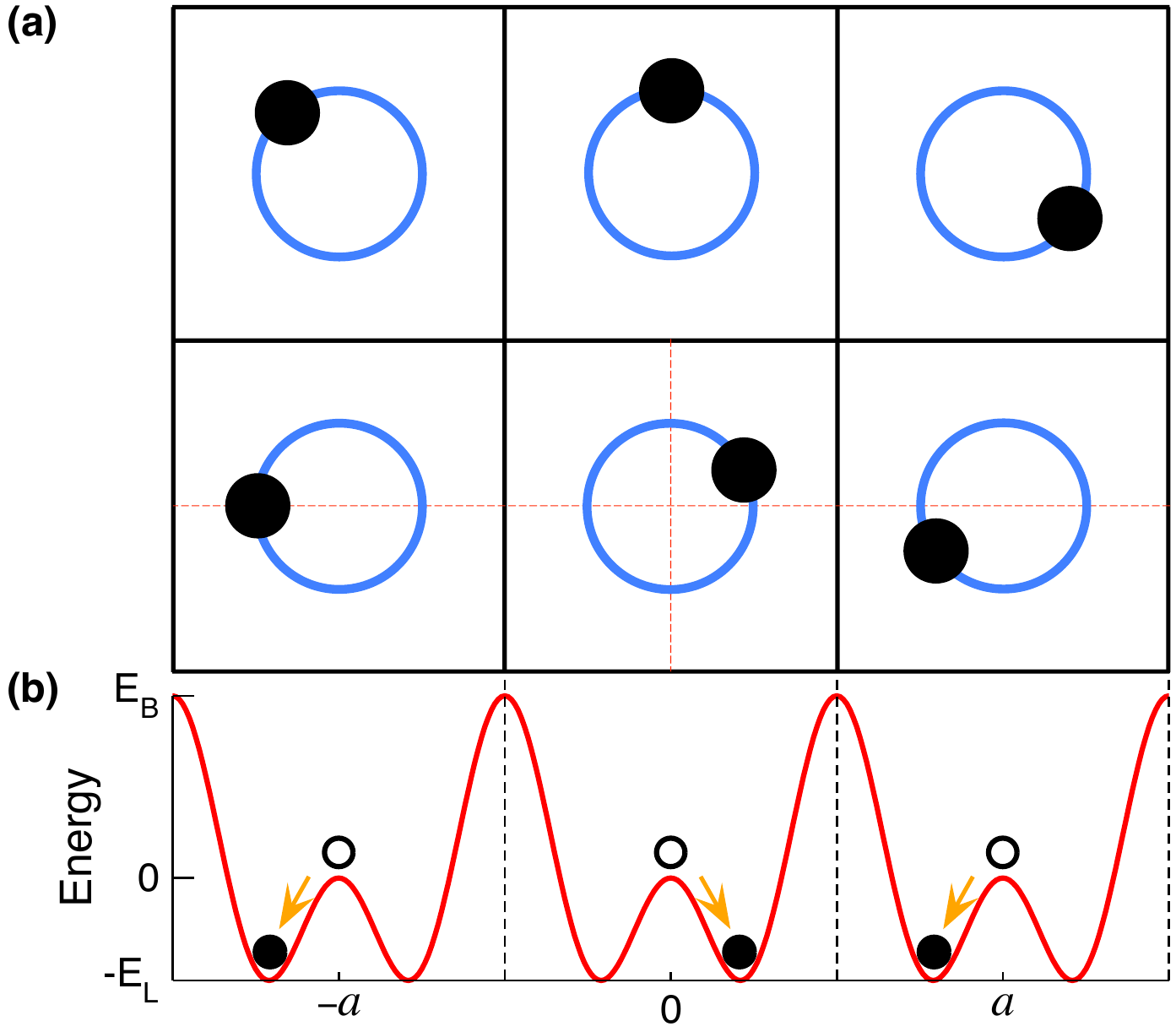}  
\vspace{-2mm}
\caption{\label{Fig2D2}(color online) 
2D square lattice with periodic sombrero-like potential, see Fig.~\ref{Fig4pot2D}.  (a) Blue circle is at PE minimum; atoms (black dots) are at arbitrary positions on this circle at low T ($kT \ll E_L$).  (b) Energy profile along the cross section [horizontal red dashed line in (a)] is the same as Fig.~\ref{Fig1ab}(b).  
}
\end{figure}

\subsection{\label{sombrero} A muffin-tin sombrero potential}
\begin{figure}[t]
\includegraphics[width=80mm]{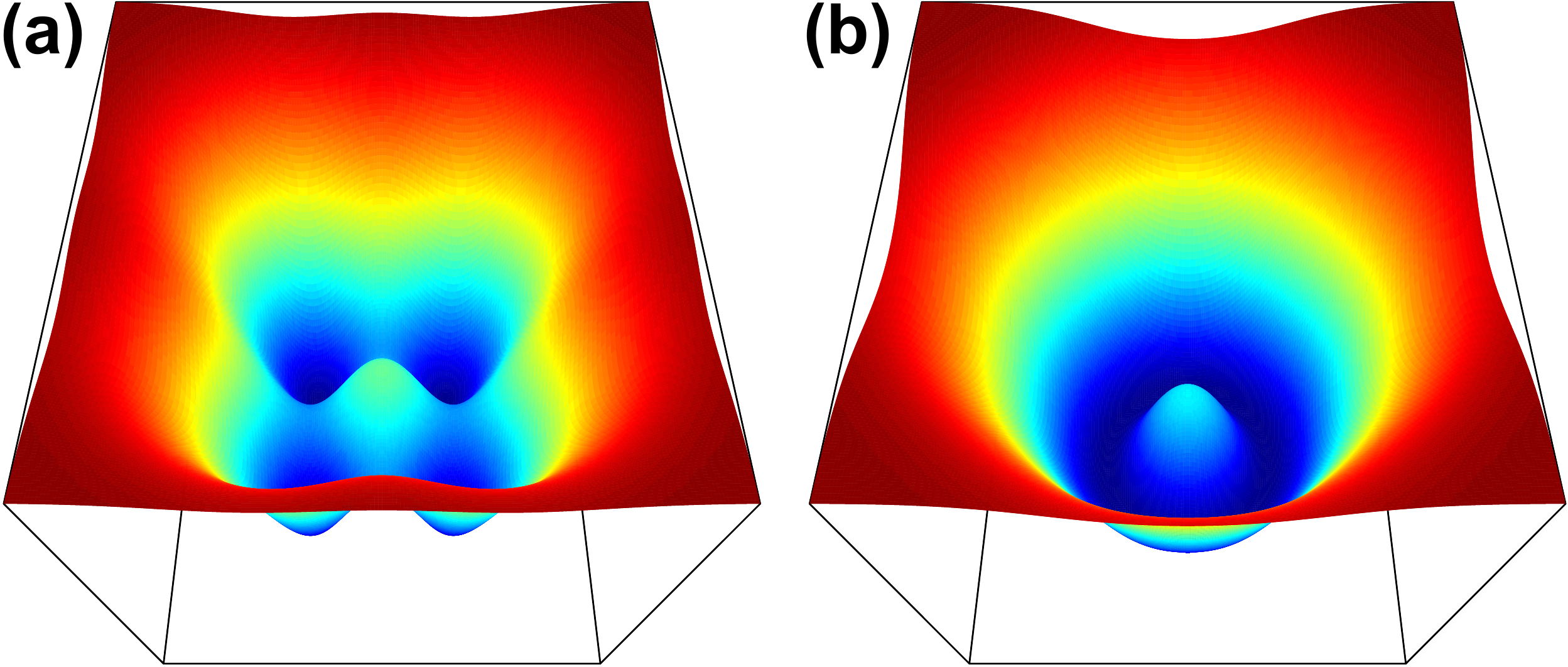}  
\vspace{-2mm}
\caption{\label{Fig5Corrugated}(color online) 
The corrugated sombrero potentials with 4 (a) and 8 (b) equidistant LPEMs, approximated by Gaussians.  
}
\end{figure}

{\par} 
	To illustrate that a MEP between LPEMs can bypass a high-symmetry crystallographic point, we show a 2D sombrero potential in Fig.~\ref{Fig4pot2D}. 
This muffin-tin type model potential is symmetric around $x=0$.  
In the radial coordinates $(r, \phi)$, it has a local energy maximum at $r=0$ and an extended energy minimum at the radius $r=x_L$ (at any angle $\phi$); 
the MEP between LPEMs is a circle $r=x_L$.  The energy barrier $E_l$ is zero for the muffin-tin sombrero potential (Fig.~\ref{Fig4pot2D}), but not for a corrugated sombrero (Fig.~\ref{Fig5Corrugated}). 
A muffin-tin type potential can form a lattice, see Fig.~\ref{Fig2D2}.
At a low temperature $kT < E_L$, atoms (black dots in Fig.~\ref{Fig2D2}) vibrate near the PE minima at $r=x_L$ (blue circles in Fig.~\ref{Fig2D2}), 
while averaged over the angles $\phi$ atomic positions happen to be at $r=0$. 
By construction, the central cross-section of the muffin-tin sombrero potential in Fig.~\ref{Fig4pot2D} is identical to the double-well potential in Fig.~\ref{Fig3doublewell}, 
and the cross-section of this periodically repeated potential in Fig.~\ref{Fig2D2} is identical to Fig.~\ref{Fig1ab}(b).

\subsection{\label{CorrugatedSombrero} Corrugated sombrero potentials}
{\par}
	Due to interatomic interactions (discussed in section~\ref{InteratomicInteractions}), an actual atomic potential differs from an idealized muffin-tin sombrero model potential, 
which approximates a corrugated sombrero with $E_l \ll E_L$ in the limit $E_l \to 0$. 
A corrugated sombrero potential has multiple local LPEMs linked by a MEP forming a closed path, with small but finite barriers $0 \le E_l \le E_L$, see Fig.~\ref{Fig5Corrugated}.   
{\par}
	At $kT \ge E_l$, atomic motion covers multiple LPEMs in a corrugated sombrero potential. However, at $kT < E_l$ an atom is trapped in the neighborhood of a single LPEM. 
Thus, there is a symmetry-breaking first-oder phase transition at $T_c$, estimated by eq.~\ref{Tc}. 
	
\begin{figure}[t]
\includegraphics[width=84mm]{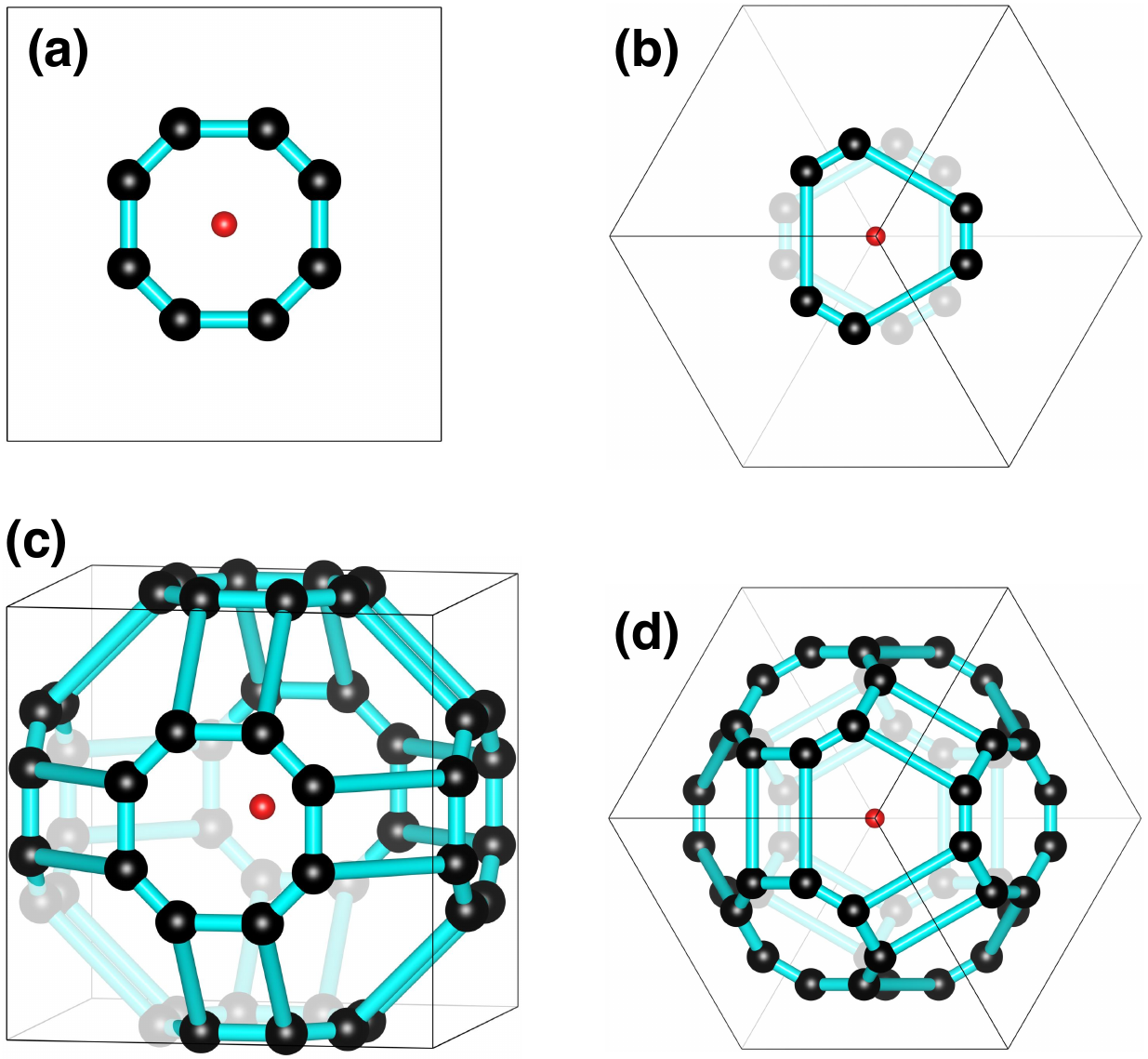}  
\vspace{-2mm}
\caption{\label{Fig48cub}(color online)
A network of the symmetry-equivalent displacements (black balls) from the center (small red dot at $x=0$) in 
(a) 2D square basin of a square lattice with 4-fold rotational symmetry;
(b) 2D hexagonal basin of a triangular lattice with 3-fold rotational symmetry (shaded is its inversion);
(c) 3D cubic lattice [shown is only the central part, not the whole cubic basin] and
(d) its [111] projection, with shading below [111] plane through 3 corners of the cube.
With a LPEM (large black dot) at each shown displacement and a local PE maximum at the center (small red dot),
the light-blue lines represent the MEP (which is not necessarily straight) between pairs of LPEMs. 
}
\end{figure}

{\par}
In Fig.~\ref{Fig48cub}, MEPs (lines) between LPEMs (dots) form a loop in 2D and a network in 3D. 
Similarity between the muffin-tin and corrugated sombreros improves with increasing number of LPEMs, see Fig.~\ref{Fig5Corrugated}.  
A cubic austenite with an unstable high-symmetry atomic position, such as NiTi, can have 48 symmetry-equivalent stable collective atomic displacements (LPEMs). 
The model sombrero potentials and a qualitative distribution of LPEMs linked by chains of MEPs in Fig.~\ref{Fig48cub} help to understand atomic motion with large athermal displacements in real materials, such as NiTi B2-type austenite.

Many solids with lattice instabilities have atomic potentials, which remind a corrugated sombrero. 
Examples include the austenitic phase of NiTi 
and the bcc phases of Ti, Zr, Hf, and Li.  

\begin{figure}[b]
\includegraphics[width=80mm]{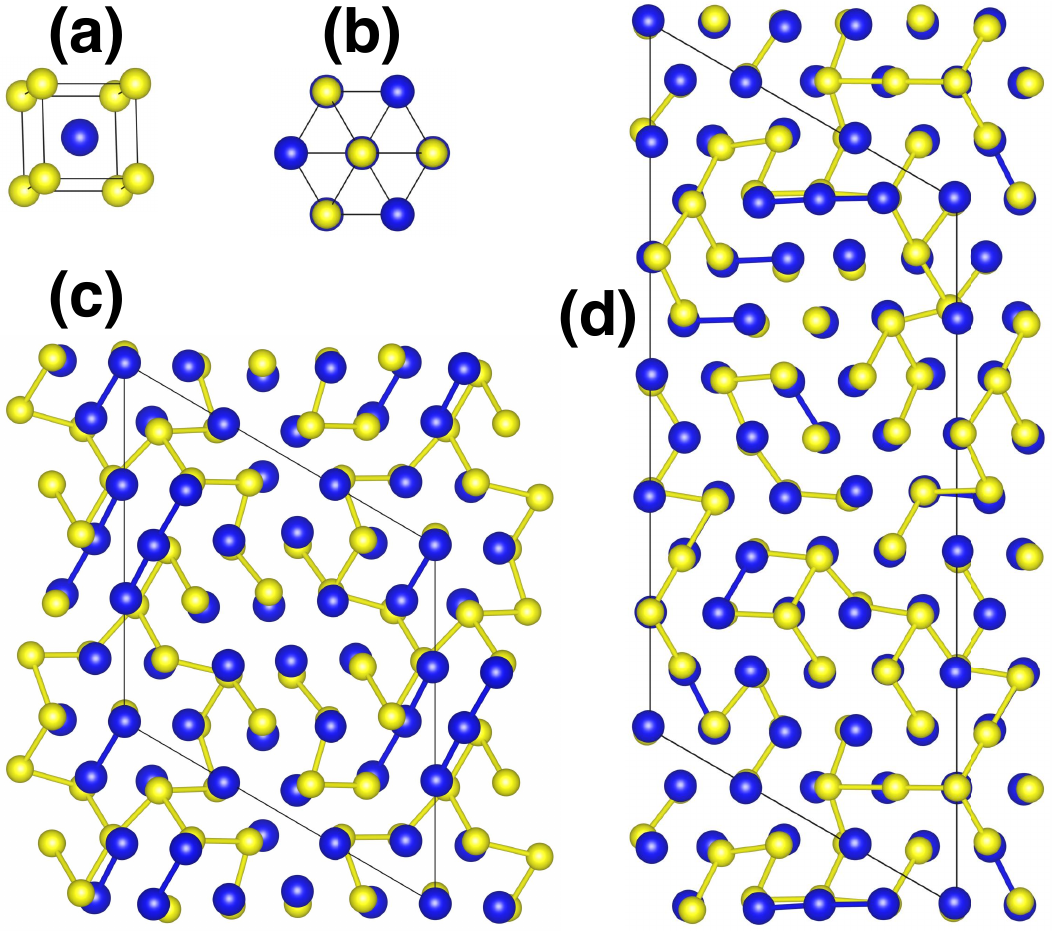}  
\vspace{-2mm}
\caption{\label{FigNi27Ti27pos}
(a) The ideal B2 structure and 
(b) its [111] projection,
with Ni (yellow) and Ti (blue) atoms; 
length of NN (Ni-Ti) bonds is 2.6~{\AA}, NNN (Ni-Ni or Ti-Ti) bonds are 3.0~{\AA}.
Stable atomic positions in cubic B2 [111] projection are shown 
in representative supercells (bounded by thin black line), 
containing 
(c) 54 atoms and (d) 108 atoms. 
The NNN Ni-Ni and Ti-Ti bonds shorter than 2.75~{\AA} are shown. 
}
\end{figure}
\begin{figure}[t]
\includegraphics[width=80mm]{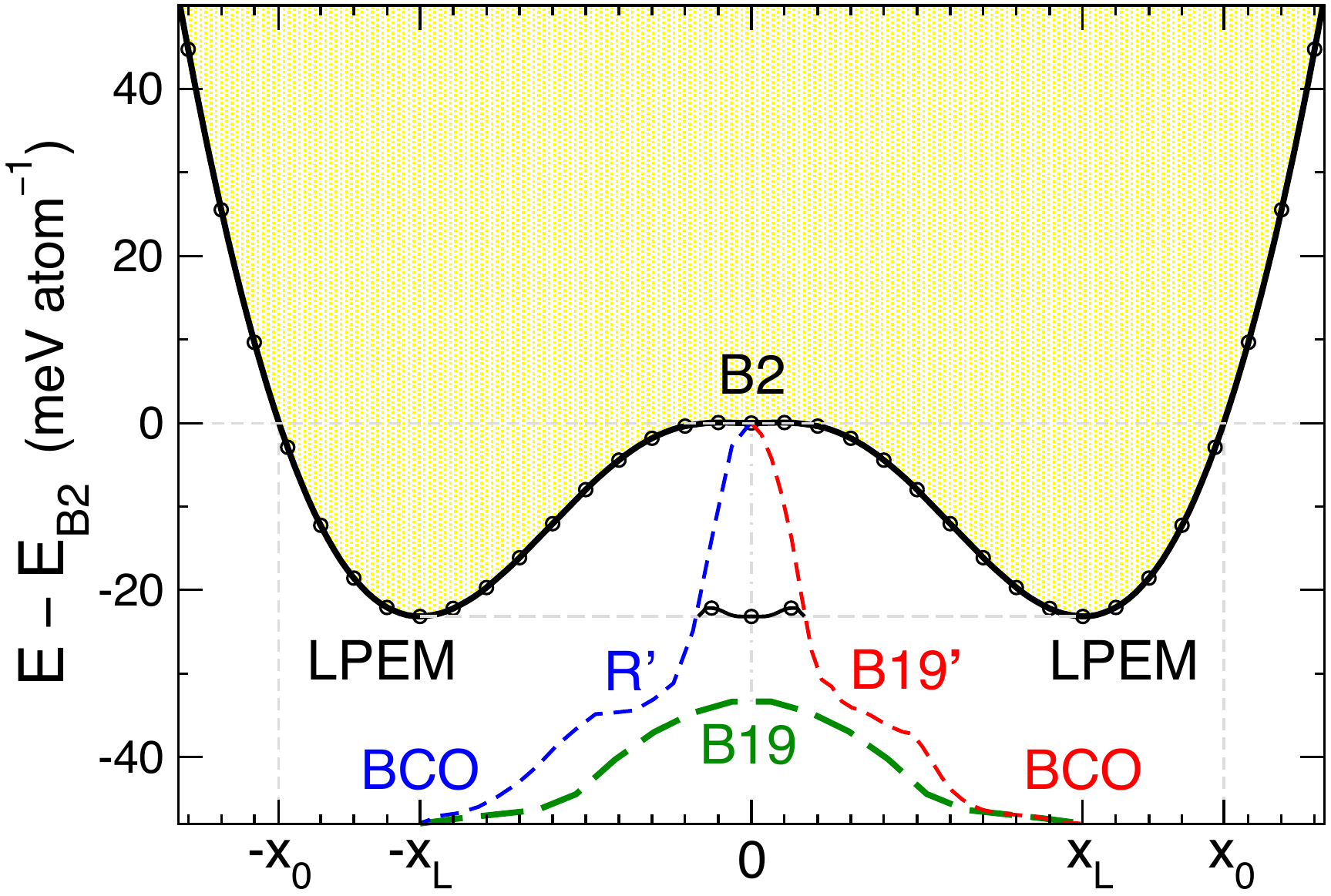}   
\vspace{-2mm}
\caption{\label{FigNiTiEx}
NiTi potential energy vs. collective atomic displacement (which changes linearly from zero at unstable B2 to $\pm x_L$ at a LPEM) 
for a MEP from B2 to a representative austenitic structure (LPEM, thick black solid line), 
from B2 to BCO ground state (via B19' or R' structures, thin dashed lines), 
and
for a transformation between two orientations of BCO martensite (via B19, thick green dashes line).  
Thin black line shows the low enthalpy barriers around an austenitic LPEM. 
}
\end{figure}

\subsection{\label{sNiTi} NiTi austenite} 
	We find that the austenitic phase of NiTi shape memory alloy above $T_c = 313\,$K
has multiple LPEMs, separated by low energy barriers $E_l \ll kT \sim E_L$.  
While B2 structure is unstable \cite{PRL113p265701}, 
a representative stable structure 
\cite{PRB90p060102} can be used for an approximate description of this solid. 

{\par}
	To obtain a representative structure, we tried several unit cells of various shapes, with various number of atoms (Fig.~\ref{FigNi27Ti27pos}). 
Computational details are in \cite{PRB90p060102,PRL113p265701} and in Appendix \ref{Acompute}.
Each supercell was heated to $800\,$K for 100~fs, cooled to $0\,$K in 800~fs (using ab initio molecular dynamics), and than fully relaxed to the nearest local potential energy minimum (using the conjugate-gradient algorithm).  Among the results, we selected the structure with the lowest energy; the smallest one had 54 atoms per representative supercell.  We checked that repeating the whole procedure with a twice larger, 108-atom unit cell (doubled along one of the 3 lattice constants) results in a different structure with the same energy per atom, see Fig.~\ref{FigNi27Ti27pos}(d).  We verified that this hexagonal 54-atom structure is indeed a local potential energy minimum with a stable phonon spectrum (Fig.~2b in \cite{PRB90p060102}), and we found that its phonon DOS compares well to that obtained from the neutron scattering experiment \cite{Fultz01}.  
Again, this representative hexagonal 54-atom structure is just an approximate representation of NiTi austenite. 
{\par}
Interatomic interactions can affect not only energy, but also atomic positions. 
Atomic displacements from the unstable ideal B2 positions are shown in Fig.~6 in \cite{PRB90p060102}  and in Fig.~\ref{FigNi27Ti27pos}~(c,d). 
The largest one ($0.66\,${\AA}) is 25\% of the nearest neighbor (NN) distance and 22\% of the B2 lattice constant ($3\,${\AA});
it is above the Lindemann criterion for melting \cite{Lindemann1910}.  
Nickel atomic radius $r(\mbox{Ni})=1.24\,${\AA} is smaller than  $r(\mbox{Ti})=1.47\,${\AA}. 
An average displacements of Ni from B2 is larger than that of Ti (Fig.~6 in \cite{PRB90p060102}).
The athermal NN-pair distribution function (Fig.~5 in \cite{PRB90p060102}) is skewed and has a substantial width (it spreads from 2.42 to 2.88 {\AA}, and from 2.48 to 2.65 {\AA} at half-maximum), in contrast to a $\delta$-function at 2.60 {\AA} for an ideal B2 single crystal.  
An average next-nearest neighbor (NNN) distance for Ni-Ni is smaller than that for Ti-Ti, while NNN Ni-Ni and Ti-Ti distances are the same in ideal B2. 
There are more short Ni-Ni bonds than short Ti-Ti bonds, see Fig.~\ref{FigNi27Ti27pos}~(c,d). 
The smallest NNN distance (for both Ni-Ni and Ti-Ti) in the austenite is shorter than the NN (Ni-Ti) distance in ideal B2 crystal. 
The NNN bonds form chains, which are linear for Ti-Ti and branching for Ni-Ni, see Fig.~\ref{FigNi27Ti27pos}~(c,d). 
A representative LPEM has a smaller energy than B2 due to optimization of interatomic distances and bond angles.  

{\par}
	The cross section of the potential energy $E$ vs. collective atomic displacement $x$ from B2 [Fig.~\ref{FigNi27Ti27pos}(a,b)] at $x\!=\!0$ 
to a representative 54-atom NiTi structure [Fig.~\ref{FigNi27Ti27pos}(c)] at $x= \pm x_L$ and beyond is a double-well $E(x)$ curve 
[see the shaded part of Fig.~\ref{FigNiTiEx}], which reminds Fig.~\ref{Fig3doublewell}. 
B2 also transforms without a barriers to the BCO ground state (see Fig.~4 in \cite{PRL113p265701}). 
The barrier (B19 in Fig.~\ref{FigNiTiEx}) for the BCO-to-BCO transformation is well below B2, see Fig.~3 in \cite{PRL113p265701} and Table~\ref{tNiTi}. 

{\par}
In general, atoms are displaced from B2 along directions, which are not high-symmetry ones, 
and a cubic structure has 48 isometries, which form the $O_h$ octahedral symmetry group, isomorphic to $S_4 \times C_2$. 
Thus, for each atom there are at least 48 LPEMs around the energy maximum at ideal B2 [Fig.~\ref{Fig48cub}(c)]. 
The barriers $E_l \sim 1\,$meV/atom between those LPEMs are quite low: 
they are comparable to the barriers between the NiTi austenite (middle of thin black line in Fig.~\ref{FigNiTiEx}) and the BCO ground state, 
which vary from only 1 to 5 meV/atom, depending on the transformation path \cite{JCP142n2p024106}.  

{\par}
Stoichiometric NiTi austenite exists at temperatures between $313\,$K and $1293\,$K:
it transforms to the low-T B19' martensite below $T_c=313\,$K (40~C, $k T_c = 27\,$meV) \cite{MT2p229y1971,Christou1972}, 
segregates above $T_s = 1293\,$K (1020~C, $kT_s = 111\,$meV), 
and melts at $T_{melt}=1586\,$K (1314~C, $k T_{melt} = 137\,$meV). 
The LPEM representing the NiTi austenitic structure is $\sim 30\,$meV/atom above the ground-state BCO and $\sim 20\,$meV/atom below unstable B2.
At $T_c<T<T_{s}$, $E_l \ll kT$, while $E_L \sim kT$.  
Thus, atoms in the NiTi austenite not only vibrate around a particular LPEM, but also move from one LPEM to another, forming a dynamically changing pattern.

\begin{figure}[t]
\includegraphics[width=80mm]{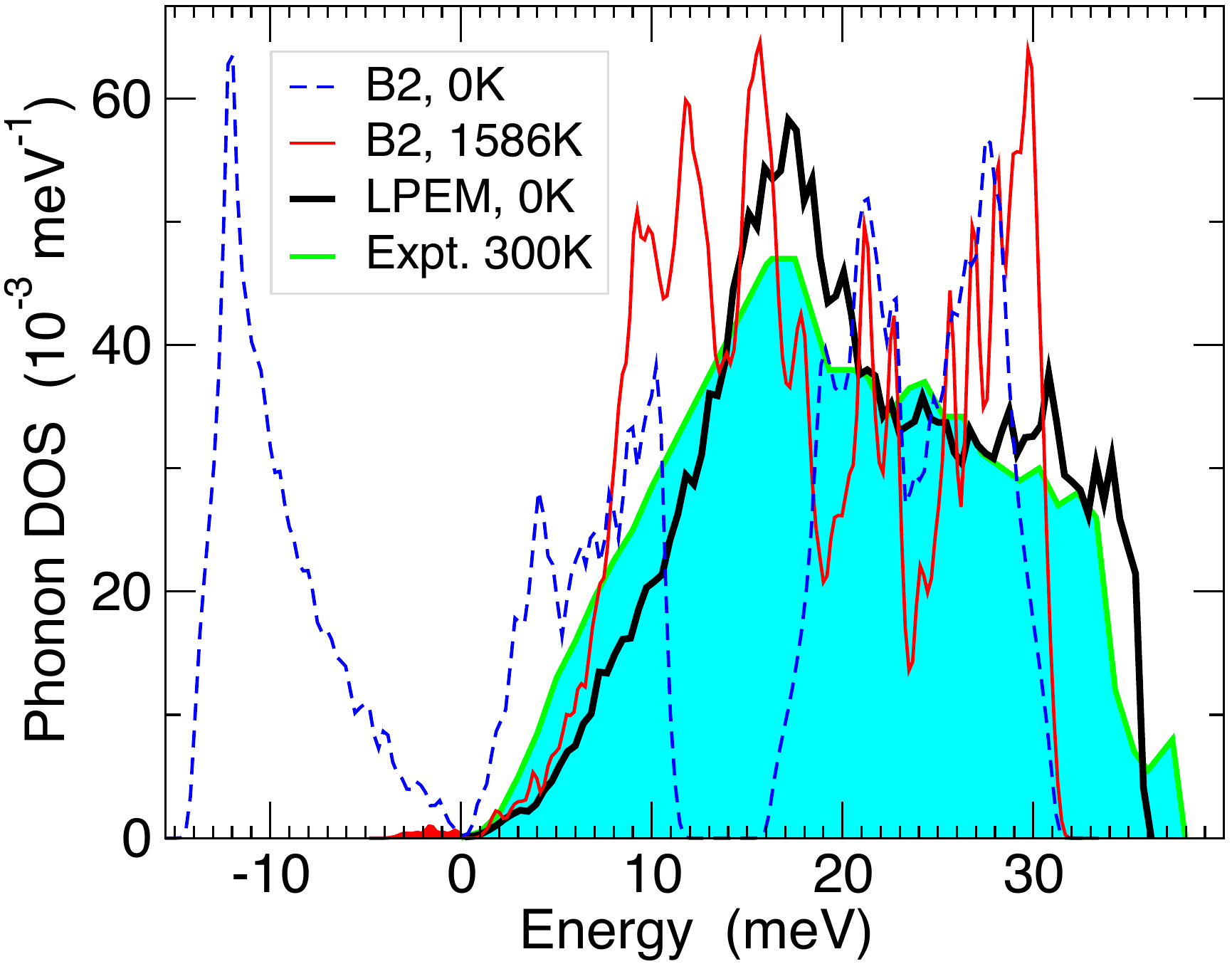}   
\vspace{-2mm}
\caption{\label{FigNiTiPhononDOS}
Phonon DOS in NiTi unstable B2 at $0\,$K (small displacements) and $1586\,$K (ThermoPhonon method \cite{ThermoPhonon}), 
and in a stable representative austenitic NiTi structure (LPEM) \cite{PRB90p060102}, 
compared to that from neutron scattering experiment \cite{Fultz01}. 
}
\end{figure}
\begin{table}[b]
\begin{tabular}{|lr|rr|rr|}
\hline
NiTi        &  \multicolumn{1}{c}{$\vartheta $} &  \multicolumn{2}{l}{$E\!-\!E_{BCO}$}  & \multicolumn{2}{l}{$E\!-\!E_{LPEM}$} \\
               &  &  meV   &  K  & meV & K \\
\hline
BCO &              $107^\circ$ &  0  &  0 &  $-29$  & $-343$ \\
B19$^\prime$ & $98^\circ$ &  8 & 93 &  $-21$ & $-250$  \\
B19 &                 $90^\circ$ &  15  &  179  &  $-14$ &  $-164$  \\
R             & &                           13  &  153  &  $-16$ &  $-190$  \\
\multicolumn{2}{l}{B2 ($E_L$)}    &  48 & 557 & +20 &  215   \\
$E_l$       & &     &                                          &	 +1  &  10  \\
\multicolumn{2}{l}{LPEM} & 29 & 343 &  0  & 0 \\
\hline
$ T_c$ & & 27  & 313 \\
$ T_s$ & & 111 & 1293 \\
\multicolumn{2}{l}{$T_{melt}$} & 137 & 1586 \\
\hline
\end{tabular}
\caption{\label{tNiTi}
Calculated energies $E$ [meV/atom] and $E/k$ [K] of NiTi structures relative to its BCO ground state  
and LPEM \cite{PRB90p060102}. 
BCO can be viewed as monoclinic B19' with angle $\vartheta = 107^\circ$.  
Energies of B19' at $\vartheta $ from $107^\circ$ to $98^\circ$ (observed in experiment \cite{JETP50p1128y1979}) vary from 0 to 8 meV/atom \cite{PRL113p265701,PRB85p014114,nmat2p307y2003}, and increase to 15.4 meV/atom for B19 at $\vartheta = 90^\circ$. 
B19 is the energy barrier for BCO-to-BCO MEP \cite{PRL113p265701}. 
The barrier $E_l$ for LPEM-to-BCO \cite{JCP142n2p024106} and LPEM-to-LPEM MEP is $\sim 1 \,$meV/atom (10~K) above LPEM.
Measured temperatures of martensitic transformation $T_c$, segregation $T_s$, and melting $T_{melt}$ \cite{MT2p229y1971,Christou1972}. 
} 
\end{table}

{\par}
	In spite of atomic motion across multiple LPEMs, 
the small displacement method applied to a stable representative austenitic NiTi structure \cite{PRB90p060102}  
provides phonon DOS, which resembles experimental one \cite{Fultz01}, see Fig.~\ref{FigNiTiPhononDOS}.  
From the other hand, increasingly large atomic displacements from B2 structure suppress the relative weight of unstable phonons.  
One can compare results of 3 phonon methods (small displacement at B2, finite displacement at B2, and small displacement at a representative LPEM)  
with the assessment based on the neutron diffraction experiment \cite{Fultz01} in Fig.~\ref{FigNiTiPhononDOS}.


\begin{figure}[b]
\includegraphics[width=80mm]{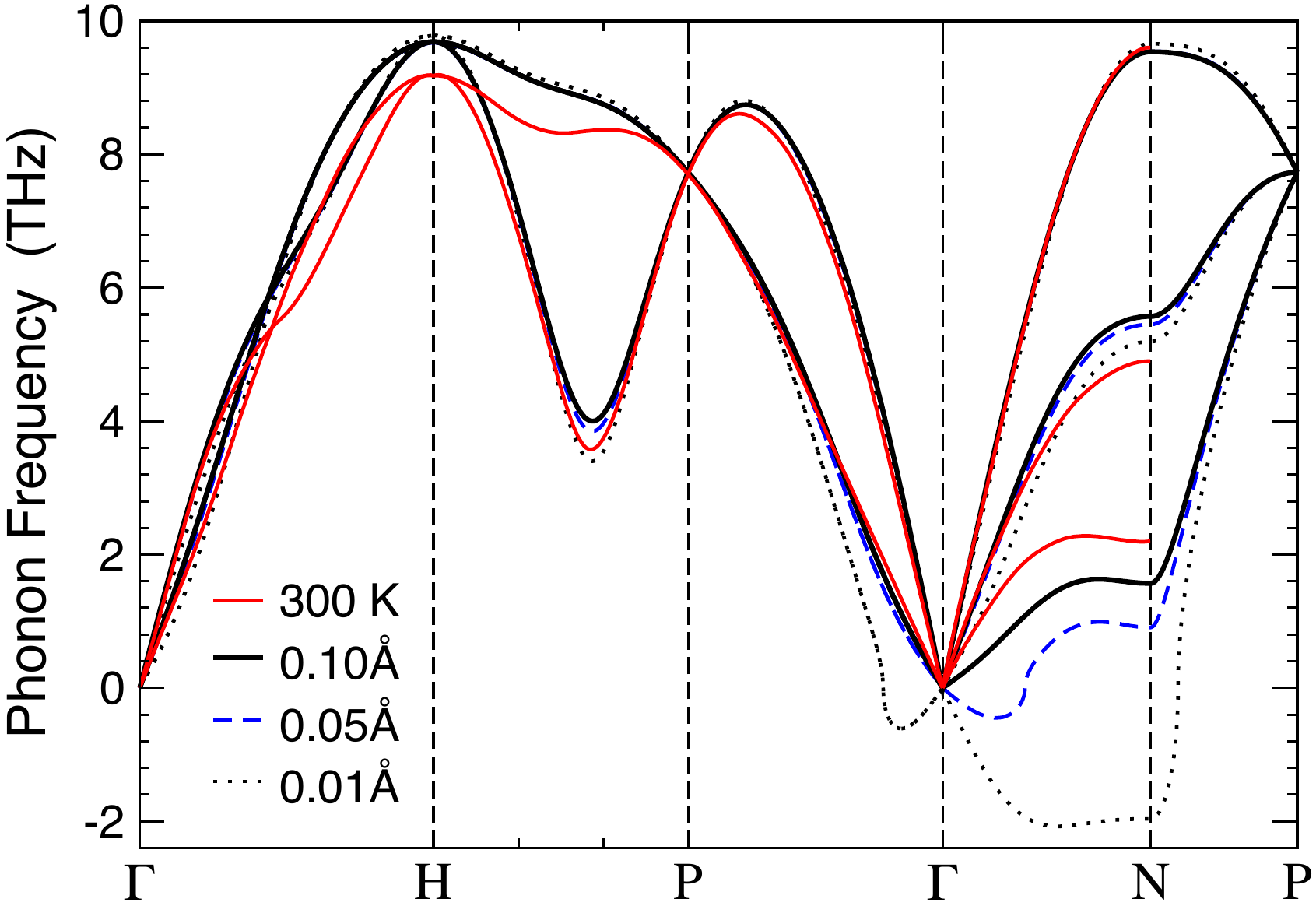}  
\vspace{-2mm}
\caption{\label{FigLiPhonon}
Phonons in bcc Li, calculated using small (0.01~\AA), medium (0.05~\AA), and large (0.1~\AA) atomic displacements and molecular dynamics (MD) at 300K \cite{PRB84p180301y2011}.
Phonon frequencies from small and medium displacement methods were rescaled to match the large displacement method at P.  Unstable phonon modes are shown as negative frequencies.
}
\end{figure}

\subsection{\label{sbccLi} Lithium}

	The bcc Li transforms to the ground-state close-packed rhombohedral 9R structure below $T_c =70\,$K \cite{PRL53p64y1984} 
and melts at $T_{melt}=453.65\,$K.  Its bcc structure is unstable.
At $T_c \ll T < T_{melt}$, including room T, 
atomic motion covers the whole central part of a PE basin, including the bcc high-symmetry point at $x=0$ and the nearby LPEMs.   
At such conditions one can use a finite atomic displacement method to calculate phonons. 

{\par} 
	We compare various methods from section~\ref{sPhonon} in Fig.~\ref{FigLiPhonon}. 
Because the bcc Li structure is unstable, the small displacement method returns unstable phonon modes around N. 
As atomic displacements become larger, the relative weight of those unstable phonons reduces, 
until they completely disappear for a sufficiently large displacement (0.1~\AA). 

{\par} 
	The molecular dynamics (MD) at a sufficiently high temperature $T \gg T_c$ provides predominantly large collective atomic displacements and a stable phonon spectrum \cite{PRB84p180301y2011}.  
The periodic boundary conditions in the $4 \times 4 \times 4$ supercell \cite{PRB84p180301y2011}, which is incommensurate with the 9R ground state, 
might  also suppress a phonon instability.  
Using MD in a smaller $3 \times 3 \times 3$ supercell, we find a very minor phonon instability at the $\Gamma$N branch, 
which reminds the result of a medium atomic displacement (0.05~\AA). 
	
{\par} 
	Overall, all 3 methods (based on the small displacements, finite displacements, and MD at a finite T) 
give phonon spectra, which are comparable everywhere, except for the $\Gamma $N branch. 
A phonon instability from the small displacement method is not a mistake, but a property of the unstable bcc Li structure. 
Suppression of those unstable phonons by using finite atomic displacements is just a computational trick.

\begin{table}[b]
\begin{tabular}{llcrrl}
\hline
\multicolumn{2}{l}{Ti}    &   &  \multicolumn{2}{c}{$E-E_{bcc}$} &  \\ 
\multicolumn{2}{c}{}      &   &   meV  &  K  \\ 
\hline
$\alpha$  & hcp  &   & $-130$ & $-1510$ \\
$\omega$ &       &   & $-128$ & $-1485$ \\
$\beta$    & bcc &   &  0 &  0  \\
\multicolumn{2}{c}{MD(1273K)}   &  & +115 & 1335 \\
\hline
\multicolumn{2}{c}{$T=1273$\,K} &  & +110 & 1273 \\
\multicolumn{2}{l}{$T_c$}         &  &  100  & 1155 \\
\multicolumn{2}{l}{$T_{melt}$} &  &   167   & 1943 \\
\hline
\end{tabular}
\caption{\label{tEbcc}
Calculated energies [meV/atom, Kelvin] 
relative to the unstable bcc Ti. 
Measures temperatures of the martensitic $\alpha$--$\beta$ transformation ($T_c$) and melting ($T_{melt}$). 
}
\end{table}

\subsection{\label{sbcc} Group 4 bcc metals: Ti, Zr, and Hf} 
\begin{figure}[t]
\includegraphics[width=80mm]{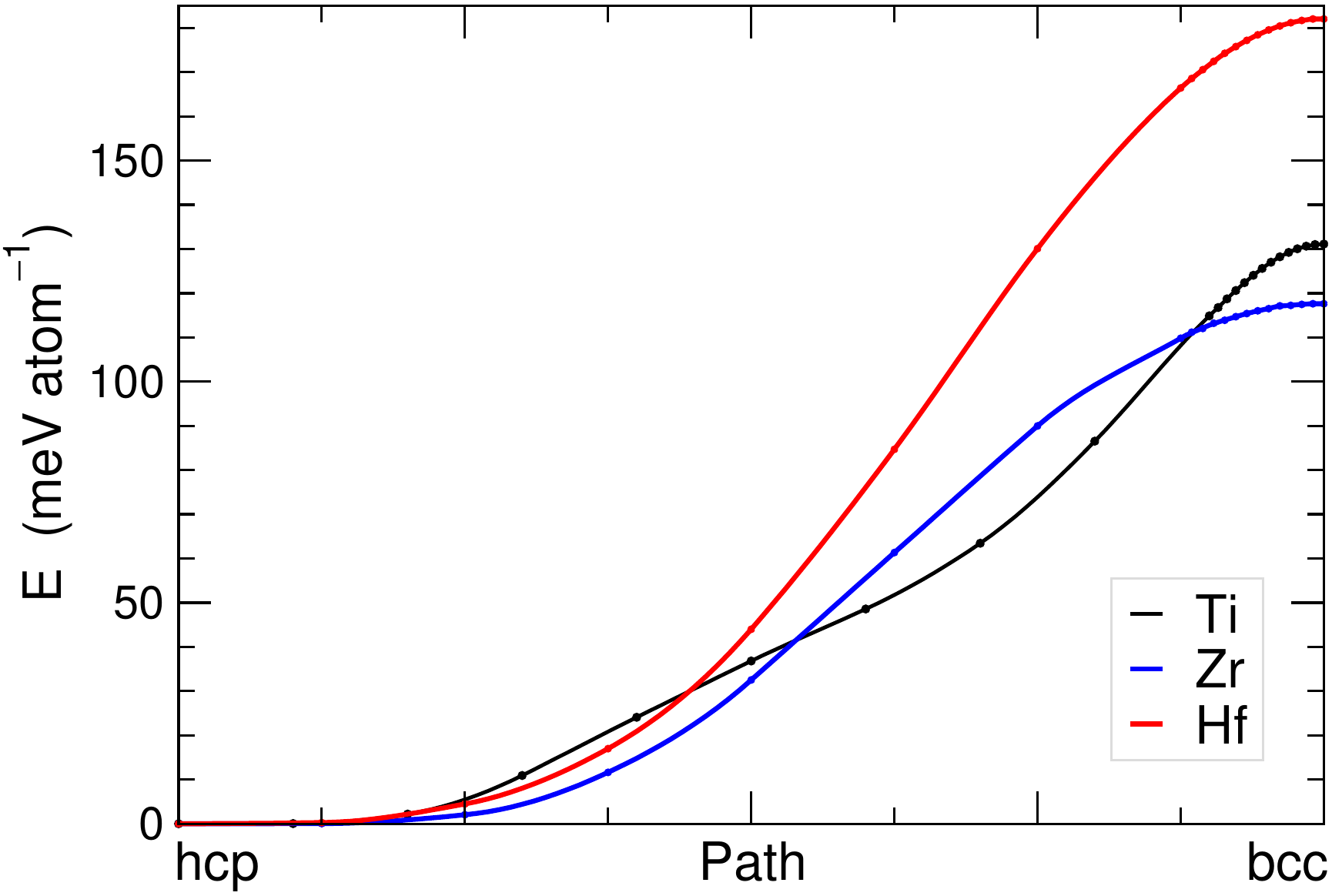}  
\vspace{-2mm}
\caption{\label{Fig_bcc2hcp_NEB}
The transformation path between hcp and bcc phases does not have an enthalpy barrier in Ti, Zr, and Hf metals. 
Their high-T bcc phase is stabilized by entropy. 
}
\end{figure}

	Materials with a lattice instability are quite common.
In particular, bcc Ti, Zr, and Hf are unstable \cite{PRL100p095901,PRB61p11221y2000,RevModPhys84p945y2012}.  
They transform from the high-T bcc $\beta$-phase to the low-T hcp $\alpha$-phase at cooling, 
and to the $\omega$-phase at pressure. 
Using the SS-NEB method, 
we find that both $\beta - \alpha$ (see Fig.~\ref{Fig_bcc2hcp_NEB}) and $\beta - \omega$ transformations are barrierless for all 3 metals, 
in agreement with the previous calculations for Ti  \cite{PRB61p11221y2000}.
Thus, ideal bcc is either a local energy maximum or a saddle point in these metals.  

{\par}
	Experiments pointed at the soft mode at $q=\frac{2}{3}(111)$, responsible for $\beta \!-\! \omega$ instability,
in Zr \cite{PRL41n25p1726y1978} and Ti \cite{PRB43p10933y1991},
and at the softening of the (110) phonon branch with temperature lowering \cite{PRB40p11425y1989}; 
the damped transverse phonons at $q=\frac{1}{2}(110)$ facilitate the $\beta \!-\! \alpha$ transition. 

{\par}
	The small atomic displacement method predicts phonon instabilities \cite{PRB61p11221y2000}
around $q=\frac{2}{3}(111)$  \cite{PRB29p1575y1984}, 
as well as in the (110) $T_1$ phonon branch (in bcc Zr  \cite{PRL58p1769y1987}). 
	The finite displacement method provided stable phonons in bcc Ti, Zr, and Hf \cite{PRB86p054119y2012}; 
similar results were obtained using a  self-consistent method with large enough atomic displacements at elevated temperature
(see Fig.~1 in \cite{PRL100p095901}, reproduced in \cite{CompMatSci44p888y2009}  and \cite{RevModPhys84p945y2012}).
	From MD at 1300K in a 128-atom $4 \times 4 \times 4$ cubic supercell (incommensurate with the $\omega$-phase),  
 a stable phonon dispersion for bcc Zr was constructed \cite{PRB84p180301y2011}.  

{\par}
	The calculated inter\-atomic force in bcc Zr for displacements  $\delta $ from bcc at 0 to $\omega$ at $x_{\omega}$ 
was found to be negative (i.e., unstable) for small  $\delta <  0.25 x_{\omega}$, but not for larger $\delta $ \cite{PRB28p6687y1983}. 
Our Fig.~\ref{Fig3doublewell} explains how the finite displacement method provides stable phonons in such cases.

{\par}
	Using MD above $T_c$, we find that 
an average PE of atoms at T of experimental bcc existence is well above that of an ideal bcc structure in Ti (Table~\ref{tEbcc}), Zr, and Hf. 
Thus, atomic motion covers the whole central part of a PE basin, including its center and all nearby LPEMs, 
which look like dimples, responsible for the lattice instability.
These PE dimples are shallow compared to $kT$, and this justifies the use of the finite displacement method for these bcc metals.  

\subsection{\label{sbccStable} Stable bcc metals: Fe, Nb} 
{\par}
Our story would be incomplete without mentioning a few conventional crystals. 
Some bcc crystals are stable. 
For example, magnetic iron has a bcc ground state \cite{PRB91p174104y2015} with a stable phonon spectrum (Fig.~1 in \cite{PRB62p273y2000}, Fig.~4 in \cite{RevModPhys73p515y2001}); its transformation to hcp has a barrier \cite{JChemPhys143n6p064707y2015}. 
The calculated phonons in bcc Nb (Fig.~2 in \cite{PRB51p6773y1995}, Fig.~3 in \cite{RevModPhys73p515y2001}) are also stable and reasonably agree with experiment \cite{JPhysCM6n31p6211y1994}. 

\subsection{\label{sTe}Tellurium}
{\par}
	The Te A8,  $\gamma$-Se, hP3, $P3_121$ trigonal structure \cite{Bradley1924} can be interpreted as a hexagonal deformation of the simple cubic (SC) structure \cite{JETP59n6p1336y1984}. 
The SC structure is unstable and transforms to A8 without a PE barrier.  
The barriers $E_l$ between various orientations of A8 are so high, that 
atoms  vibrate around a single LPEM, which is a stable A8 crystal, observed in experiment \cite{ActaCrystC45p941y1989}. 

{\par}
	Tellurium is a provocative example, which can be interpreted either as a stable Te A8 structure with a single LPEM (see section~\ref{s1Dharmonic}),
or as a highly unstable SC structure with multiple LPEMs (section~\ref{s1Danharmonic}), separated by high barriers $E_l$ ($kT < E_l < E_L$, section~\ref{sGeneric}).

\section{\label{sDiscussion} Discussion}
{\par} 
	In general, a dynamically polymorphic solid has a higher lattice entropy than a conventional crystal. 
Entropy is proportional to the logarithm of the number of states, 
and vibration around a single LPEM has fewer states than atomic motion across multiple LPEMs (see Figs.~\ref{Fig1ab}(b) and \ref{Fig48cub}). 
Many solid phases are stabilized by entropy at a finite T,  
and dynamic polymorphism is very common in those high-temperature solid phases. 

{\par} 
	Examples of dynamically polymorphic phases include 
solids with lattice instabilities (such as bcc Li, B2 AFM FeRh, and NiTi austenite),
crystals with a mobile interstitial dopant (some of metal hydrides and boron steels),
polymers and organic molecules with rotating molecular units, 
and numerous solid phases, dynamically stabilized by entropy at a finite T.

\section{\label{sConclusion}Summary}
{\par} 
	Solids with dimpled atomic potentials are quite ubiquitous among natural and industrial materials.  Some of them have multiple local minima of the atomic potential energy.
They include many high-temperature solid phases, which  
 have a higher lattice entropy than conventional crystals. 
Examples include many anharmonic crystals with lattice instabilities, such as bcc Li, Ti, Zr, and Hf elemental solids, B2-type antiferromagnetic FeRh and NiTi austenite. 
To understand properties of these materials, we have constructed simplified models.   
We compared three methods for calculating phonons, 
applied them to conventional harmonic crystals and solids with a lattice instability, 
and discussed various types of atomic motion. 
We predicted a first-order phase transition in cooled dynamically polymorphic phases, 
and provided an estimate of the transition temperature.


\acknowledgments
{\bf Acknowledgments: }
We acknowledge Dr. Vitalij Pecharsky, Viktor Balema, Yaroslav Mudryk,  Andrey Smirnov,  Dmitry Mendelev, Klaus Ruedenberg (Ames, USA),  Brent Fultz (CalTech, USA), 
Dario Alf\'e (London, UK), and Igor Abrikosov (Link\"opings, Sweden)
 for discussion.
Development of novel methods was supported by the U.S. Department of Energy (DOE), Office of Science, Basic Energy Sciences, Materials Science and Engineering Division. 
Applications to caloric materials are supported by the U.S. DOE,  Advanced Manufacturing Office of the Office of Energy Efficiency and Renewable Energy through 
CaloriCool\textsuperscript{TM} -- the Caloric Materials Consortium established as a part of the U.S. DOE Energy Materials Network.  
The research was performed at the Ames Laboratory, which is operated for the U.S. DOE by Iowa State University under contract DE-AC02-07CH11358.

\appendix
\section{\label{Acompute}Computational details}
{\par}
	DFT calculations were performed using the plane-wave pseudopotential-based VASP code \cite{VASP1,VASP2}. 
We used the generalized gradient approximation (GGA) \cite{GGA}, and 
 a projected augmented wave (PAW) basis \cite{PRB50p17953}, 
with convergence obtained by a second Broyden’s method \cite{PRB38p12807y1988}. 
We used the high accuracy  \cite{VASP1,VASP2}  with 
default values of the plane-wave energy cutoff and augmentation charge cutoff (e.g., 337~eV and 544.6 eV for NiTi \cite{PRB90p060102,PRL113p265701}). 
The total energies and forces were calculated using $k$-meshes with at least 50 $k$-points per {\AA}$^{-1}$. 

{\par}
	The solid-state nudged elastic band (SS-NEB) method \cite{GSSNEB} with up to two climbing images \cite{JCP142n2p024106} 
was combined with DFT \cite{VASP1,VASP2} to address transformations \cite{PRL113p265701}.  
To obtain result in Fig.~\ref{FigNiTiEx}, we linearly extrapolated atomic coordinates from ideal B2 to the austenitic structure \cite{PRB90p060102} and beyond
in a representative 54-atom supercell  (Fig.~\ref{FigNi27Ti27pos}c). We remind, that a linear extrapolation is not necessarily the MEP, 
thus there could be a fictitious barrier in Fig.~\ref{FigNiTiEx}, if the MEP was twisted \cite{JCP142n2p024106}.

{\par}
	The group 4 metals (Ti, Zr, Hf) were addressed using DFT+U \cite{LDAUTYPE2} with $(U\!-\!J)=2.2$\,eV \cite{PRB93p020104}.  
The endpoint bcc and hcp structures were fully relaxed.  Next, we used the SS-NEB method \cite{GSSNEB} to find the minimal enthalpy path. 
High precision of DFT calculations \cite{VASP1,VASP2} was achieved with the plane-wave energy (augmentation charge) cutoff of 
223.0 (328.9)  eV for Ti, 
193.3 (243.2) eV for Zr, and
275.5 (335.1) eV for Hf.

{\par}
	Li and LiH were addressed by combining VASP \cite{VASP1,VASP2},  ThermoPhonon  \cite{ThermoPhonon}, and Phonopy \cite{Phonopy} codes.
The small displacement method was applied using the Phon code \cite{Phon}.

\bibliography{Network}
\end{document}